\documentclass[iop]{emulateapj-rtx4}
\usepackage{color}
\usepackage{graphicx}
\usepackage{tabularx}
   \newcolumntype{C}{>{\centering\arraybackslash}X}
   \newcolumntype{L}{>{\raggedright\arraybackslash}X}
   \newcolumntype{R}{>{\raggedleft\arraybackslash}X}

\newcommand{\Ka}{K$\alpha$}

\newcommand{\NH}{$N_{\rm H}$}
\newcommand{\Msun}{$M_{\odot}$}
\newcommand{\Mch}{$M_{\rm Ch}$}

\newcommand{\kTe}{$kT_{\rm e}$}

\shorttitle{Double Ring Structure in N103B}
\shortauthors{Yamaguchi et al.}

\begin{document}

\title{Discovery of Double-Ring Structure in The Supernova Remnant N103B: \\
Evidence for Bipolar Winds from a Type I\lowercase{a} Supernova Progenitor}

\author{
Hiroya Yamaguchi\altaffilmark{1,2},
Fabio Acero\altaffilmark{3},
Chuan-Jui Li\altaffilmark{4},
and 
You-Hua Chu\altaffilmark{4}
}
\email{yamaguchi@astro.isas.jaxa.jp}

\altaffiltext{1}{Institute of Space and Astronautical Science, JAXA, 3-1-1 Yoshinodai, Sagamihara, 
	Kanagawa 229-8510, Japan}
\altaffiltext{2}{Department of Physics, The University of Tokyo, Hongo, Bunkyo, Tokyo, 113-0033, Japan}
\altaffiltext{3}{
D\'epartement d'Astrophysique-AIM, CEA/DRF/IRFU, CNRS/INSU, Universit\'e Paris-Saclay, Universit\'e de Paris,
Gif-sur-Yvette, France}
\altaffiltext{4}{Institute of Astronomy and Astrophysics, Academia Sinica, No.1, Sec. 4, Roosevelt Rd., Taipei 10617, Taiwan}




\begin{abstract}

The geometric structure of supernova remnants (SNR) provides a clue to unveiling the 
 pre-explosion evolution of their progenitors. 
Here we present an X-ray study of N103B (0509--68.7), a Type Ia SNR in the Large 
Magellanic Cloud, that is known to be interacting with dense circumstellar matter (CSM). 
Applying our novel method for feature extraction to deep Chandra observations, 
we have successfully resolved the CSM, Fe-rich ejecta, and intermediate-mass element 
(IME) ejecta components, and revealed each of their spatial distribution. 
Remarkably, the IME ejecta component exhibits a double-ring structure, 
implying that the SNR expands into an hourglass-shape cavity 
and thus forms bipolar bubbles of the ejecta. 
This interpretation is supported by more quantitative spectroscopy that 
reveals a clear bimodality in the distribution of the ionization state of the IME ejecta.
These observational results can be naturally explained if the progenitor binary system 
had formed a dense CSM torus on the orbital plane prior to the explosion, providing 
further evidence that the SNR N103B originates from a single-degenerate progenitor. 

\end{abstract}




\section{Introduction}
\label{sec:intro}

Type Ia supernovae (SNe Ia) are utilized as distance indicators to constrain 
the cosmological parameters and the nature of the dark energy 
\citep[e.g.,][and references therein]{Suzuki12}. 
Although it is generally accepted that SNe Ia originate from a thermonuclear explosion 
of a carbon-oxygen white dwarf (WD) in a binary system, their evolution channel is 
still under debate. One hypothesis is that the thermonuclear explosion is caused 
by a merger of two WDs, which is called the double degenerate (DD) scenario 
\citep[e.g.,][]{Iben84,Webbink84} and often associated with an explosion of 
a sub-Chandrasekhar-mass (sub-\Mch) WD \citep[e.g.,][]{Fink10,Shen18}. 
Another hypothesis is the so-called single degenerate (SD) scenario, 
where a WD accretes hydrogen-rich matter from a non-degenerate companion and 
explodes when the WD mass approaches \Mch\ \citep[e.g.,][]{Whelan73,Nomoto82}. 
Recent observations tend to support the DD scenario, 
based on the lack of a surviving companion in Type Ia supernova remnants 
\citep[SNRs Ia: e.g.,][]{Schaefer12,Kerzendorf13,Kerzendorf18,Ruiz18} 
or the lack of the signature of interaction between SN Ia ejecta 
and a non-degenerate companion in either early-phase lightcurve 
\citep[e.g.,][]{Brown12,Olling15} or nebula-phase spectra 
\citep[e.g.,][]{Shappee13a,Lundqvist15,Sand18,Dimitriadis19,Tucker19}.
Moreover, the delay time distribution of SNe Ia favors the DD channel 
as the major contributor to these events \citep[e.g.,][]{Totani08,Maoz10}.
However, contributions of the near-\Mch\ progenitors 
are also necessary from the perspective of the cosmic chemical evolution, 
suggesting that both SD and DD channels may end up as SNe Ia
\citep[e.g.,][]{Seitenzahl13b,Hitomi17b,Kobayashi20}.
It is, therefore, crucial to identify the exact relationship between the 
different evolution channels and different observational characteristics 
of SNe Ia (and their remnants), which in turn enables 
even more robust distance measurements in cosmology.

An important prediction of the SD scenarios is the presence of mass outflow 
from the binary system during the pre-SN evolution \citep[e.g.,][]{Hachisu96}. 
Therefore, detection of circumstellar matter (CSM) from SNe Ia is 
naturally interpreted as evidence of their SD progenitor origin 
\citep[e.g.,][]{Hamuy03,Simon09,Sternberg11,Silverman13}.
The presence of CSM is confirmed also in a handful of young SNRs Ia, 
which offer a unique opportunity to investigate their CSM's spatial distribution, 
providing the key for understanding how their progenitor systems have evolved exactly. 
One of such objects is Kepler's SNR, the relic of SN\,1604 \citep{Blair91,Gerardy01}. 
In this SNR, a significant amount of the CSM is located at the northern rim and 
the central region, with little evidence for dense materials in the south 
\citep{Blair07,Reynolds07,Williams12}. 
The north-south asymmetry of the CSM is suggested to be formed by 
the wind activity of the progenitor system moving toward the north 
\citep{Borkowski92,Chiotellis12}. 
The other CSM concentration found near the projected SNR center is 
claimed as evidence for a disk distribution of the CSM 
around the pre-explosion progenitor system \citep{Burkey13}. 

\newpage

The SNR N103B (0509--68.7), located near the edge of the central bar of 
the Large Magellanic Cloud (LMC), is another SNR Ia that is known to be 
interacting with a dense CSM \citep[e.g.,][]{Williams14}. 
The age of the SNR is estimated to be less than 1000~years from the 
light echo observations and shock velocity measurement 
\citep{Rest05,Ghavamian17,Williams18}.
Because of its proximity to the ionized superbubble around the 
NGC\,1850 cluster, 
N103B was initially identified as a core-collapse SNR \citep{Chu88}. 
This interpretation was supported by X-ray observations using the 
XMM-Newton Reflection Grating Spectrometer (RGS), based on the detection 
of strong O emission \citep{Heyden02}. 
However, a high-resolution imaging study and spatially-resolved spectroscopy 
with Chandra revealed that the O emission likely originated from the CSM 
rather than the SN ejecta \citep{Lewis03}. 
The authors also argued that the estimated ejecta mass of Si and Fe 
were consistent with predictions of typical SN Ia products, 
confirming the previous spectroscopic results of ASCA observations \citep{Hughes95}.

Similar to Kepler's SNR, N103B exhibits a highly asymmetric morphology 
in various wavelengths, with the brightness enhanced toward the west 
\citep[e.g.,][]{Williams99,Lewis03,Alsaberi19}. 
Using high-resolution optical images from the Hubble Space Telescope (HST), 
\cite{Li17} spatially resolved the complex radiative clumps that are 
located inside an Balmer-dominated filamentary shell. 
The density of the clumps is generally higher than 100\,cm$^{-3}$ 
and even reaches $\sim$\,5000\,cm$^{-3}$ in the densest regions, 
providing clear evidence of their CSM origin. 
The H$\alpha$ emission from these clumps is significantly redshifted, 
suggesting the one-sided distribution of the CSM \citep{Li17,Ghavamian17}. 
This asymmetry is thought to have originated from a proper motion of 
the progenitor binary system toward the west.
It is notable, however, that radio observations with the Australia Telescope 
Compact Array (ATCA) and Atacama Large Millimeter/submillimeter Array (ALMA) 
discovered a giant molecular cloud toward the southeast of the SNR \citep{Sano18}. 
The cloud shows an expanding gas motion with its spatial extent 
along the SNR rim, which strongly suggests the SNR-cloud interaction 
indeed taking place there.
The average H$_2$ density in the interacting region is estimated to be 
$\sim$\,1500\,cm$^{-3}$, even higher than the typical atomic hydrogen density in the west. 
This fact suggests that the absence of the H$\alpha$ emission in the eastern side 
of the SNR is not due to the low ambient density, 
but is due to the chemical composition dominated by the molecular hydrogen. 
The elemental abundance of the CSM is consistent with the local interstellar 
medium (ISM) of the LMC \citep{Blair20}, implying that the companion of 
N103B must have been a main-sequence star, rather than more evolved ones, 
such as an asymptotic giant branch star\footnote{In contrast, two surviving 
companion candidates found in other SNRs Ia in the LMC are 
both located in the red giant branch \citep{Li19}.}$\!$. 
To summarize, N103B is an SNR Ia likely originating from a relatively young 
SD progenitor system that has experienced substantial mass loss prior to 
the explosion, and is now interacting with the CSM with a spherically 
asymmetric distribution. 
These characteristics make this object an ideal laboratory for probing 
into the nature of SD progenitors in general.

This paper presents an X-ray study of N103B based on the latest deep 
Chandra observations. Although the observations were conducted primarily 
for an expansion velocity measurement \citep{Williams18}, the data are 
remarkably suitable for spatially-resolved spectroscopy as well. 
Our immediate aim is to determine the geometry of the SN ejecta and CSM, 
independently from the previous multi-wavelength studies,
in order to constrain the pre-explosion activity of its progenitor. 
To achieve this goal, we apply the new feature extraction technique 
established by \cite{Picquenot19}, and perform detailed spectral analysis. 
The uncertainties quoted in the text and table and the error bars given in 
the figures represent a 1$\sigma$ confidence level, unless otherwise stated.


\section{Analysis and Results}
\label{sec:analysis}

\subsection{Observations and Data Reprocess}
\label{ssec:reduction}

The SNR N103B was observed by Chandra in the Spring of 2017 (between March 20 and June 1) 
using the ACIS-S3 chip primarily for the expansion velocity measurement \citep{Williams18}.
The resulting data consist of 12 separate observations with different satellite 
roll angles. Since the calibration database (CALDB) for the ACIS instrument 
has been updated after the observations, we reprocess the data using CIAO 
version 4.12.1 and the latest CALDB (v.4.9.2.1) for the present work. 
After the standard data screening, we obtain a total effective exposure 
of 393\,ks.

\subsection{Component Extraction: Methods}
\label{ssec:comp_m}

We conduct the feature extraction based on 
the Generalized Morphological Components Analysis (GMCA) described in 
\cite{Picquenot19}\footnote{This technique was first introduced by 
\cite{Bobin16} for the cosmic microwave background (CMB) reconstruction.} 
to disentangle different physical components present in the data and 
to identify the extraction regions of interest. 
Imaging spectroscopy instruments, such as the Chandra ACIS, provide 
`three-dimensional' data cubes of the photon positions and energy $(x, y, E)$. 
The GMCA is a source separation method that looks for clusters of voxels in the data 
cube with similar spectral signatures. This method assumes 
that different physical components (e.g., shocked ejecta, nonthermal emission) 
have different morphologies in addition to different spectral signatures, 
and fully exploits the multi-dimensional aspect of the X-ray data. 
It is a blind approach and has no prior instrumental or physical 
information. Therefore, the components do not come with a label, and physical 
interpretation is to be made by users based on the decomposed spectra and images. 
Inputs to the algorithm are the data cube and the user-defined number $N$ of 
components to retrieve. The output is a set of $N$ images and spectra with 
different morphological and spectral signatures.

\begin{figure*}[t!]
  \begin{center}
	\includegraphics[width=16.8cm]{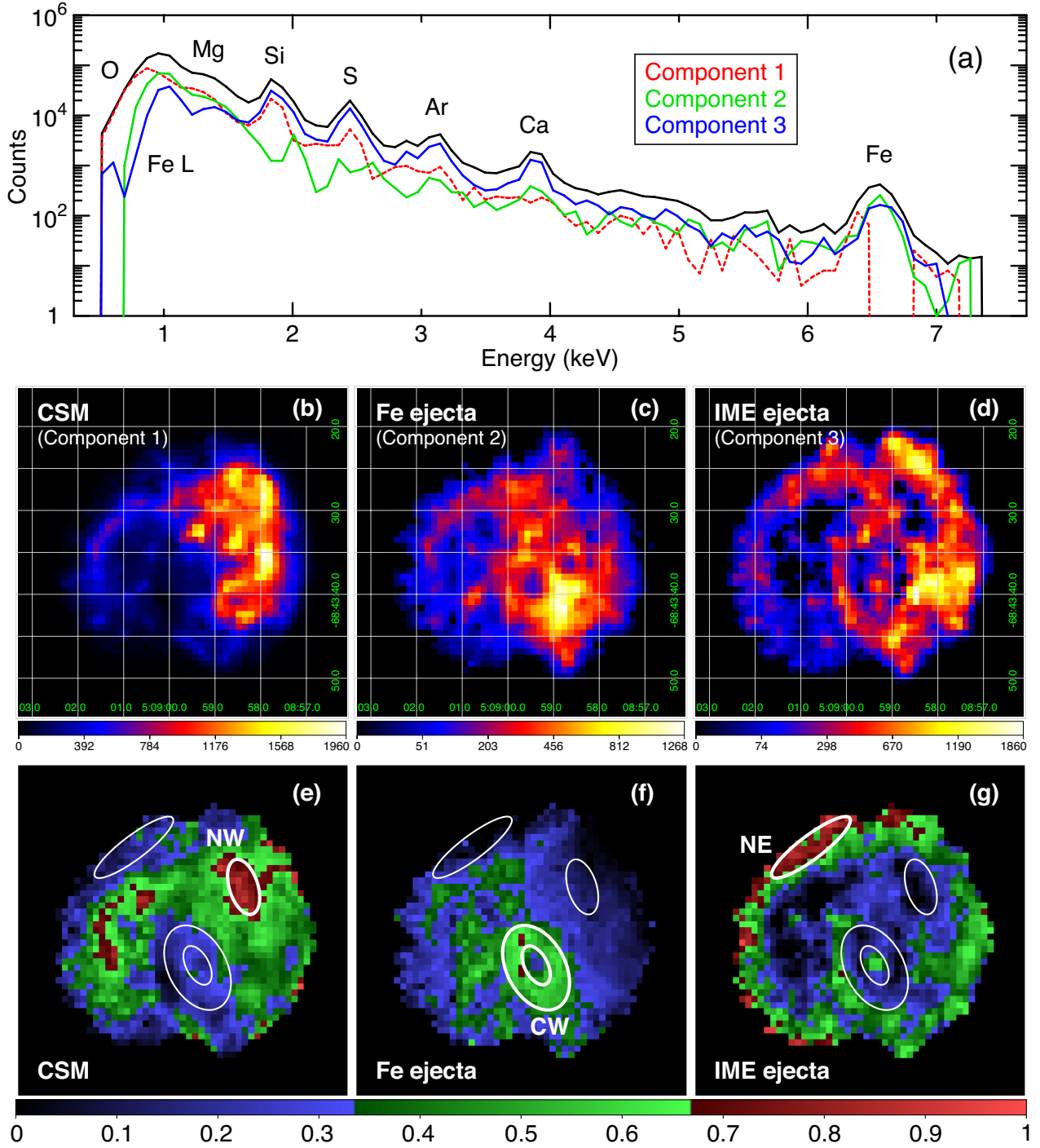}
	\caption{\footnotesize
		Results of the GMCA applied to the SNR N103B.\ 
		{\it Top}: Retrieved spectra of the three components.\ 
		{\it Middle}: Spatial distribution of the retrieved components. 
		The pixel values correspond to the photon counts normalized by the count 
		in the single brightest pixel. Panel (b) is given in the linear scale, 
		whereas Panels (c) and (d) are in the square root scale.\ 
		{\it Bottom}: Fraction of each component (see text) given in the linear scale. 
		The ellipses indicate where the spectra shown in Figures~\ref{spec_comp} 
		and \ref{all_spectra} are extracted. 
	\label{image_comp}}
  \end{center}
\end{figure*}

To build the data cube needed for our feature extraction applied to N103B, we merge 
the event lists from the 12 separate observations. 
The spatial and spectral binning are optimized to be $0''\!.75$ and 87.6\,eV 
(corresponding to six spectral channels of the ACIS), respectively, 
to achieve an ideal balance between the statistics in each voxel and 
retainment of enough spatial and spectral information. 
The number of components is fixed to three, obtaining the set of decomposed 
images and spectra given in Figure~1. 
An addition of the fourth component results in overfitting of the data;
two of the four components show similar images and spectra. 
Therefore, we conclude that the optimal number for our data is three.

\subsection{Component Extraction: Results}
\label{ssec:comp_r}

\begin{figure*}[t!]
  \begin{center}
          \vspace{2mm}
	\includegraphics[width=17cm]{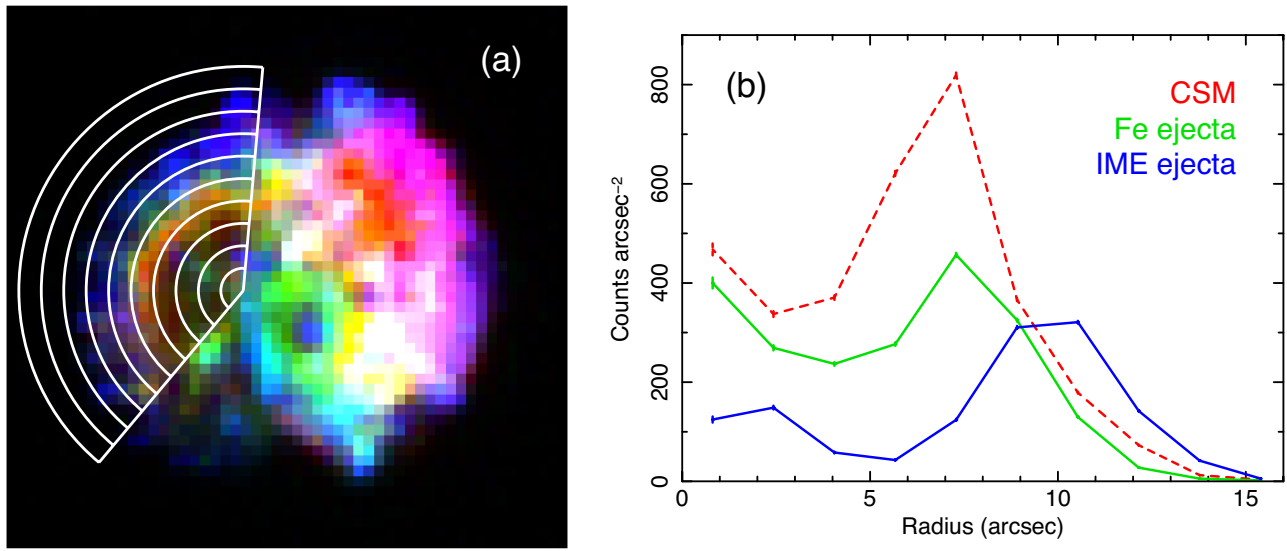}
	\caption{\footnotesize
		(a) Three-color image of the SNR N103B. Red (dashed), Green, and blue are 
		the CSM, Fe ejecta, and IME ejecta components, respectively. 
		(b) Radial profiles extracted from the sectors shown in Panel (a), 
		which confirms that the IME ejecta component peaks at a larger radius 
		than the others.
			\label{profile}}
  \end{center}
\end{figure*}

The first panel (Figure~\ref{image_comp}a) shows the spectrum of each retrieved component. 
Component~1 (red) is characterized by the strong O K emission below 0.7\,keV, 
suggesting a CSM origin of this components. 
Component~2 (green) consists mainly of the Fe L-shell ($\sim$\,1\,keV) and 
K-shell ($\sim$\,6.5\,keV) emission, and thus can be associated with the 
Fe-rich ejecta originating from the nuclear statistical equilibrium 
achieved during the progenitor's explosion. 
The retrieved spectrum of Component~3 (blue) dominates the line flux of 
the intermediate-mass elements (IME: Si, S, Ar, Ca) and reproduces also a part 
of the Fe emission detected in the spectrum of the entire SNR. 
This characteristic is consistent with the emission from incomplete Si burning 
products typically observed in other SNRs Ia (e.g., Tycho, Kepler). 
Given the results, we hereafter call the Components 1, 2, and 3 as 
``CSM'', ``Fe ejecta'', and ``IME ejecta'', respectively. 
This identification will be verified later with detailed analysis of 
the X-ray spectra as well as comparison with optical observations.

Figures\,\ref{image_comp}b, \ref{image_comp}c, and \ref{image_comp}d present 
the spatial distribution of the CSM, Fe ejecta, and IME ejecta, respectively. 
The pixel value indicates the photon counts of each component. 
The CSM image exhibits several clumpy features in the west, 
which spatially coincide with the optical nebula knots 
\citep[][see also \S\ref{sec:interpretation}]{Li17,Ghavamian17}. 
The emission from the Fe ejecta is strongest in the southwest, 
consistent with the previous XMM-Newton observation \citep{Heyden02} 
but is revealed more clearly with a higher spatial resolution. 
The IME ejecta are also prominent in the west, but its morphology is 
distinctly different from those of the other components. 
There are several remarkable features revealed in this image. 
First, not only in the bright western rim but also in the fainter east, 
the spatial extent of this component is larger than those of the other components, 
which is also confirmed in the radial profiles given in Figure\,\ref{profile}.
Second, its morphology does not seem to be a single shell but looks apparently 
a double ring structure, {\it like `pretzel'}, consisting of two elliptical 
shells crossing each other around the geometrical center of the SNR. 

The third row of Figure~\ref{image_comp} shows the fraction of 
each component, defined as $f_i = N_i/(N_1+N_2+N_3)$, 
where $N_i$ is the photon counts of Component $i$ normalized by 
the maximum pixel count obtained for each component. 
The enhancement of the IME ejecta at the SNR rim is confirmed even more 
clearly in Figure~\ref{image_comp}g. It should be noted that a similar radial 
structure was revealed in the equivalent width maps of the He-like Si and S 
emission by the previous Chandra study \citep{Lewis03}, where little 
azimuthal dependence in the equivalent width was found as well. 
Our analysis confirms the same trend with an independent approach 
applied to the much longer exposure data.

\begin{figure*}[t!]
  \begin{center}
	\includegraphics[width=17cm]{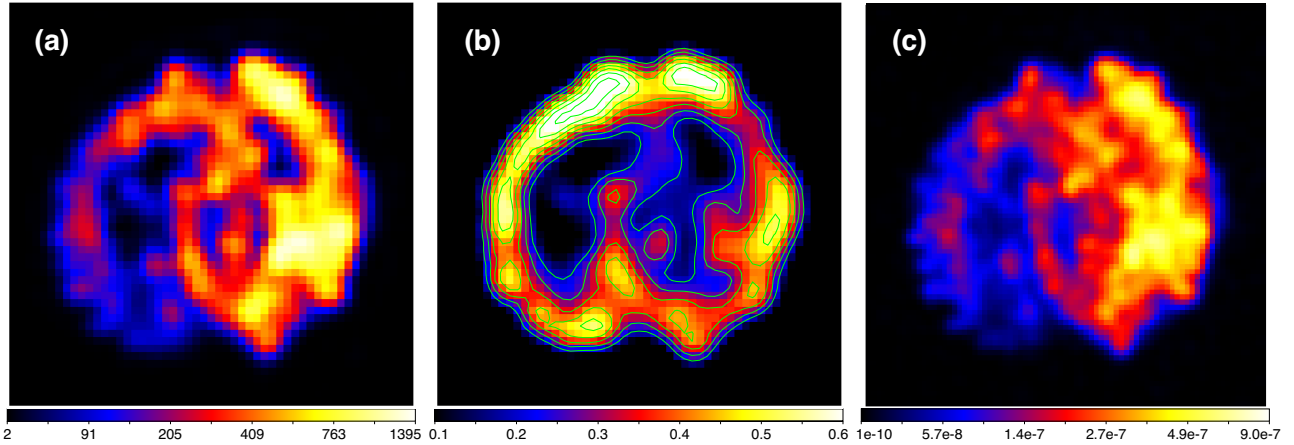}
	\caption{\footnotesize
		Spatial distribution of the IME ejecta component. 
		Panels (a) and (b) are essentially the same of Figures~1d and 1g, 
		but smoothed to highlight the double-ring structure.\ 
		Panel (c) is a simple flux image of the Si K band (1.80--2.05\,keV) 
		in the unit of photon\,cm$^{-2}$\,s$^{-1}$. 
	\label{ime_image}}
  \end{center}
\end{figure*}

The double-ring structure of the IME ejecta component is more clearly seen 
in Figures~\ref{ime_image}a and \ref{ime_image}b, which are essentially 
the same as Figures~\ref{image_comp}d and \ref{image_comp}g, respectively, 
but are smoothed to emphasize the image contrast. 
We also show in Figure~\ref{ime_image}c a narrow band flux image of 
the Si K band (1.80--2.05\,keV) generated using the {\tt flux\_obs} 
script in the CIAO package. The double ring shape already appears in 
this simply-processed image, indicating that this intriguing structure 
is not an artifact owing to our new analysis methods.

\subsection{Spectral Characteristics}
\label{ssec:spec1}

\begin{figure}[t!]
  \begin{center}
	\includegraphics[width=7.8cm]{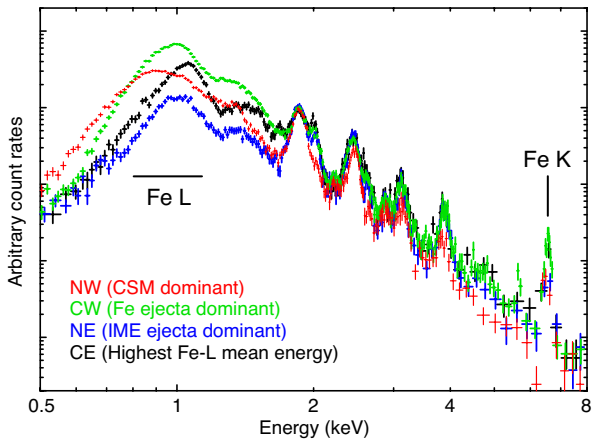}
	\caption{\footnotesize
		Comparison of the ACIS spectra extracted from the regions indicated in 
		Figure~1 or 4: NW (red), CW (green), NE (blue), and CE (black).
	\label{spec_comp}}
  \end{center}
\end{figure}

\begin{figure}[t!]
  \begin{center}
	\includegraphics[width=7.2cm]{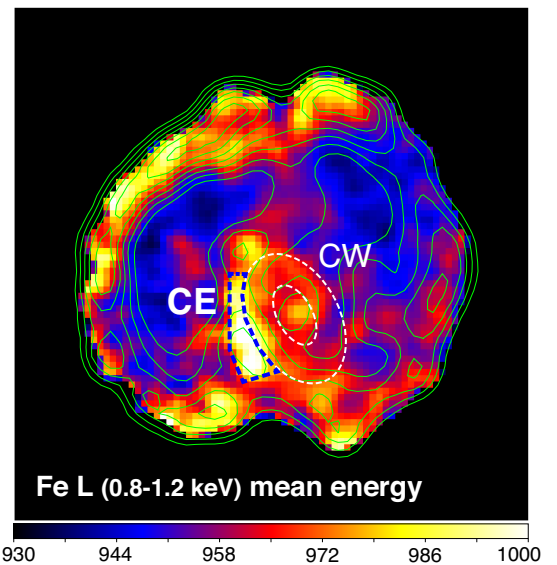}
	\caption{\footnotesize
		Mean energy of the Fe L-shell emission in 0.8--1.2\,keV.
		The contours are the same as those presented in Figure~\ref{ime_image}b.
	\label{mean_fe}}
  \end{center}
\end{figure}

\begin{figure}[t!]
  \begin{center}
	\includegraphics[width=7.8cm]{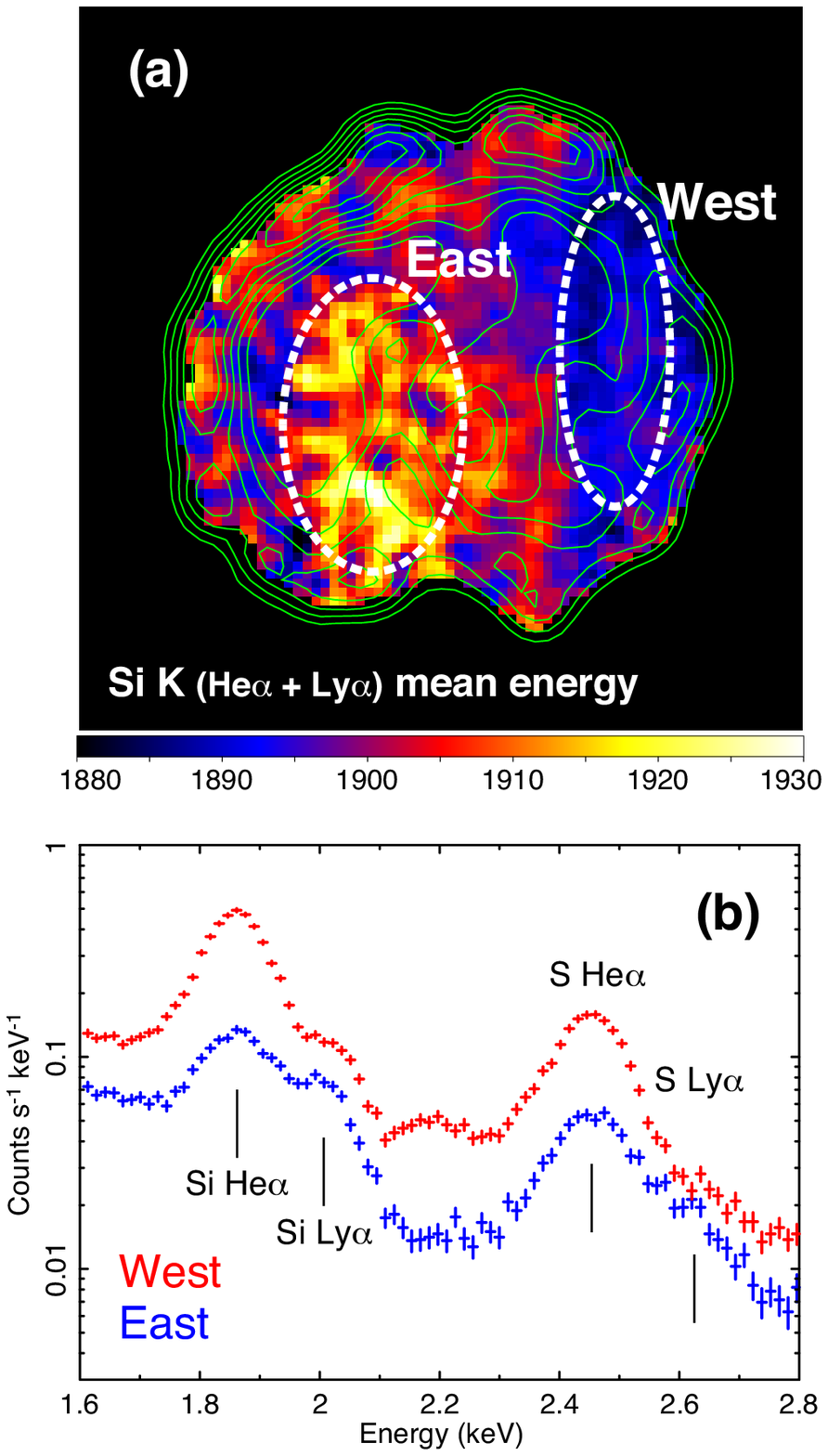}
	\caption{\footnotesize
		(a) Mean energy of the Si K-shell emission in 1.80-2.05\,keV.
		The contours are the same as those presented in Figure~\ref{ime_image}b. \ 
		(b) the 1.6--2.8-keV spectra extracted from the regions indicated in Panel (a).
	\label{mean_si}}
  \end{center}
\end{figure}

We extract spectra from three representative regions, NW, CW, and NE in 
Figure~\ref{image_comp}, containing pixels with large $f_i$ values for each 
of the identified components. The background data are taken from the nearby 
source-free region and subtracted from the sources. The obtained spectra 
are compared in Figure~\ref{spec_comp}. 
As expected, the spectrum of NW (red) exhibits an enhanced flux below $\sim$\,0.7\,keV, 
and that of CW (green) shows the strongest Fe emission (both L-shell and K-shell) 
with respect to the IME emission. 
The centroid energy of the Fe K emission in the CW spectrum is found to be 
$\sim$\,6.54\,keV, consistent with the Suzaku measurement of 
the entire SNR's spectrum \citep{Yamaguchi14b}.

Also remarkable in Figure~\ref{spec_comp} is that the shape of the Fe L-shell 
blend around 1\,keV is substantially different among the spectra, implying that 
the dominant charge states of Fe ions are different from region to region. 
To investigate this variety more comprehensively, we map the mean energy of the 
X-ray photons detected in the 0.8--1.2\,keV band using the {\tt mean\_energy\_map} 
script in CIAO. The result is shown as the color map in Figure~\ref{mean_fe}, 
where the $f_3$ image (same as Figure~\ref{ime_image}b) is overplotted in contours. 
We find that the mean energy of the Fe L emission is well correlated with 
the $f_3$ value, suggesting that Fe ions in the IME component are generally 
more highly ionized. In contrast, the regions dominated by the CSM component 
have lower mean energies (i.e., lower charge state of Fe). 
Notably, the highest mean is found at the region labeled ``CE'' (blue dashed polygon) 
in Figure~\ref{mean_fe}, which corresponds to the eastern side of the ``west ring''. 
The spectrum of this region is also shown in Figure~\ref{spec_comp} with the black 
data points, confirming the Fe L blend with a significantly high centroid.

Figure~\ref{spec_comp} also indicates considerable spatial variability in the 
Si Ly$\alpha$/He$\alpha$ ratio. We thus generate another mean energy map using 
the 1.80--2.05\,keV photons in Figure~\ref{mean_si}a. 
Interestingly, the mean energy distribution is clearly bimodal; the interior 
of the east ring (hereafter East bubble) globally has the high mean energies, 
and that of the west ring (West bubble) the lower means. The Si K band 
spectra of the East and West bubbles are shown in Figure~\ref{mean_si}b. 
The Ly$\alpha$/He$\alpha$ flux ratios are indeed largely different between 
the two regions, indicating that the IME ejecta are more highly ionized 
in the East bubble than in the West one.

\subsection{Spectral Modeling}
\label{ssec:spec2}

Here we analyze all the spectra given in Figure~\ref{spec_comp} more quantitatively. 
In this section, the data from the 12 separate observations are combined only 
for the display purpose but left unmerged in the actual spectral analysis. 
We instead fit 12 individual spectra from the identical sky regions simultaneously, 
using the response matrices generated independently.

Since our GMCA method has identified the physical components likely associated with CSM, 
Fe ejecta, and IME ejecta, we initiate spectral modeling with three components of absorbed 
thermal plasmas with different compositions. The free parameters are listed in Table~1. 
For the foreground extinction, we consider both Galactic and Magellanic absorption columns 
by introducing a {\tt TBabs} model with the solar abundance \citep{Wilms00} and 
a {\tt TBvarabs} model with the LMC abundance \citep{Dopita19}. The hydrogen 
column density \NH\ of the former is fixed at $6 \times 10^{20}$\,cm$^{-2}$ 
\citep{Dickey90}, whereas that of the latter is left as a free parameter. 
For the CSM component, we introduce a {\tt vapec} model \citep{Foster12} 
with the elemental abundances fixed at the LMC mean values, 
assuming a collisional ionization equilibrium (CIE) for this plasma. 
Since the previous work suggested non-equilibrium ionization (NEI) 
for the plasma corresponding to this component \citep{Heyden02,Lewis03}, 
we also try to apply a {\tt vnei} model alternatively. 
However, the ionization parameter $\tau = \int n_{\rm e} \, dt$ is always 
obtained to be $\gtrsim$\,$10^{12}$\,cm$^{-3}$\,s (in all the four regions), 
consistent with the CIE. 
For the Fe ejecta component, we assume a pure-metal plasma in the NEI condition, 
with no admixture of elements other than Fe and Ni\footnote{A model based 
on this assumption is often applied to Fe ejecta in SNRs Ia, such as 
Tycho SNR \citep{Yamaguchi17}.}$\!$. The abundance ratio between 
these two elements are assumed to be the solar value of \cite{Wilms00}. 
Since no hydrogen is contained in this component, its emission measure is 
defined as a product of the electron and Fe densities and the emitting volume, 
$n_{\rm e}n_{\rm Fe}V$, instead of $n_{\rm e}n_{\rm H}V$. 
Lastly, another NEI model is applied to the IME ejecta component, 
where abundances of the elements not listed in Table~1 are all set to zero. 
The parameter $v$, the line-of-sight velocity, is allowed to vary to 
reproduce a small red- or blueshift detected in the IME emission. 
The spectral fitting is performed based on the $C$-statistic \citep{Cash79} 
on unbinned spectra using the XSPEC software version 12.10.0c.

\begin{table*}[p]
\begin{center}
\caption{Best-fit spectral parameters.
  \label{tab}}
  \begin{tabular}{llcccc}
\hline \hline
Component & Parameter & NW & CW & NE & CE \\
\hline
Absorption & \NH$^{\rm LMC}$ ($10^{21}$ cm$^{-2}$) & $3.3 \pm 0.3$ & $3.5 \pm 0.1$ & $6.5_{-1.1}^{+1.0}$ & $2.0_{-0.7}^{+0.4}$ \\
CSM & $kT_{\rm e}$ (keV) & $0.29 \pm 0.01$ & $0.72 \pm 0.01$ & $0.20 \pm 0.01$ & $0.64_{-0.05}^{+0.06}$ \\
~ & $n_{\rm e} n_{\rm H} V$ ($10^{58}$ cm$^{-3}$) & $5.0_{-0.6}^{+0.5}$ & $4.0 \pm 0.1$ & $2.9_{-1.1}^{+1.5}$ & $0.16_{-0.01}^{+0.04}$ \\
Fe ejecta & $kT_{\rm e}$ (keV) & $8.6_{-0.4}^{+0.5}$ & $7.1_{-0.5}^{+0.4}$ & $11_{-3}^{+4}$ & $42 \pm 30$ \\
~ & $\tau$ ($10^{10}$ cm$^{-3}$\,s) & $1.3 \pm 0.1$ & $3.4 \pm 0.1$ & $2.3_{-0.2}^{+0.3}$ & 3.4 (fixed) \\
~ & $n_{\rm e} n_{\rm Fe} V$ ($10^{53}$ cm$^{-3}$) & $1.4 \pm 0.1$ & $7.1_{-0.1}^{+0.5}$ & $0.68_{-0.12}^{+0.13}$ & $0.26 \pm 0.07$ \\
IME ejecta & $kT_{\rm e}$ (keV) & $1.6 \pm 0.1$ & $1.7 \pm 0.1$ & $2.2 \pm 0.4$ & $1.6_{-0.1}^{+0.2}$ \\ 
~ & Mg (solar) & $0.77_{-0.12}^{+0.09}$ & $1.7 \pm 0.1$ & $2.2 \pm 0.4$ & $1.6_{-0.1}^{+0.2}$ \\
~ & Si (solar) & $3.6_{-0.4}^{+0.3}$ & $6.0_{-0.4}^{+0.2}$ & $9.7_{-2.0}^{+1.9}$ & $4.3_{-0.5}^{+0.6}$ \\
~ & S (solar)  & $3.4 \pm 0.3$ & $7.2_{-0.5}^{+0.3}$ & $9.2_{-1.6}^{+1.5}$ & $4.4 \pm 0.5$ \\
~ & Ar (solar) & $3.6 \pm 0.6$ & $9.3_{-0.8}^{+1.0}$ & $8.4 \pm 1.6$ & $5.5_{-0.7}^{+1.1}$ \\
~ & Ca (solar) & $4.5_{-1.0}^{+1.2}$ & $17 \pm 2$ & $17 \pm 3$ & $6.4 \pm 1.4$ \\
~ & Cr, Mn (solar) & (tied to Ca) & $61_{-11}^{+14}$ & (tied to Ca) & (tied to Ca) \\
~ & Fe, Ni (solar) & $0.98_{-0.12}^{+0.08}$ & $2.1_{-0.1}^{+0.2}$ & $1.8_{-0.5}^{+0.4}$ & $2.7 \pm 0.3$ \\
~ & $\tau$ ($10^{10}$ cm$^{-3}$\,s) & $7.5_{-0.5}^{+0.8}$ & $26_{-4}^{+7}$ & $9.3_{-1.9}^{+3.0}$ & $37_{-10}^{+7}$ \\
~ & $v$ (km\,s$^{-1}$) & $-1400_{-100}^{+200}$ & $95_{-47}^{+53}$ & $-740_{-50}^{+90}$ & $1200 \pm 100$ \\
~ & $n_{\rm e} n_{\rm H} V$ ($10^{57}$ cm$^{-3}$) & $6.2_{-0.5}^{+0.8}$ & $5.7_{-0.9}^{+0.1}$ & $2.3_{-0.4}^{+0.9}$ & $3.7_{-0.6}^{+0.4}$ \\
\hline
c-stat/d.o.f. & ~ & 3956/6126 & 5010/6125 & 3914/6126 & 3744/6127 \\
\multicolumn{2}{l}{Major component} & CSM & Fe ejecta & IME ejecta & IME ejecta \\
\hline
\end{tabular}
\tablecomments{
The abundances are relative to the solar values of \cite{Wilms00}. \\
The negative values of the radial velocity ($v$) indicates blueshift. \\
}
\end{center}
\end{table*}

\begin{figure*}[p]
  \begin{center}
	\includegraphics[width=16.5cm]{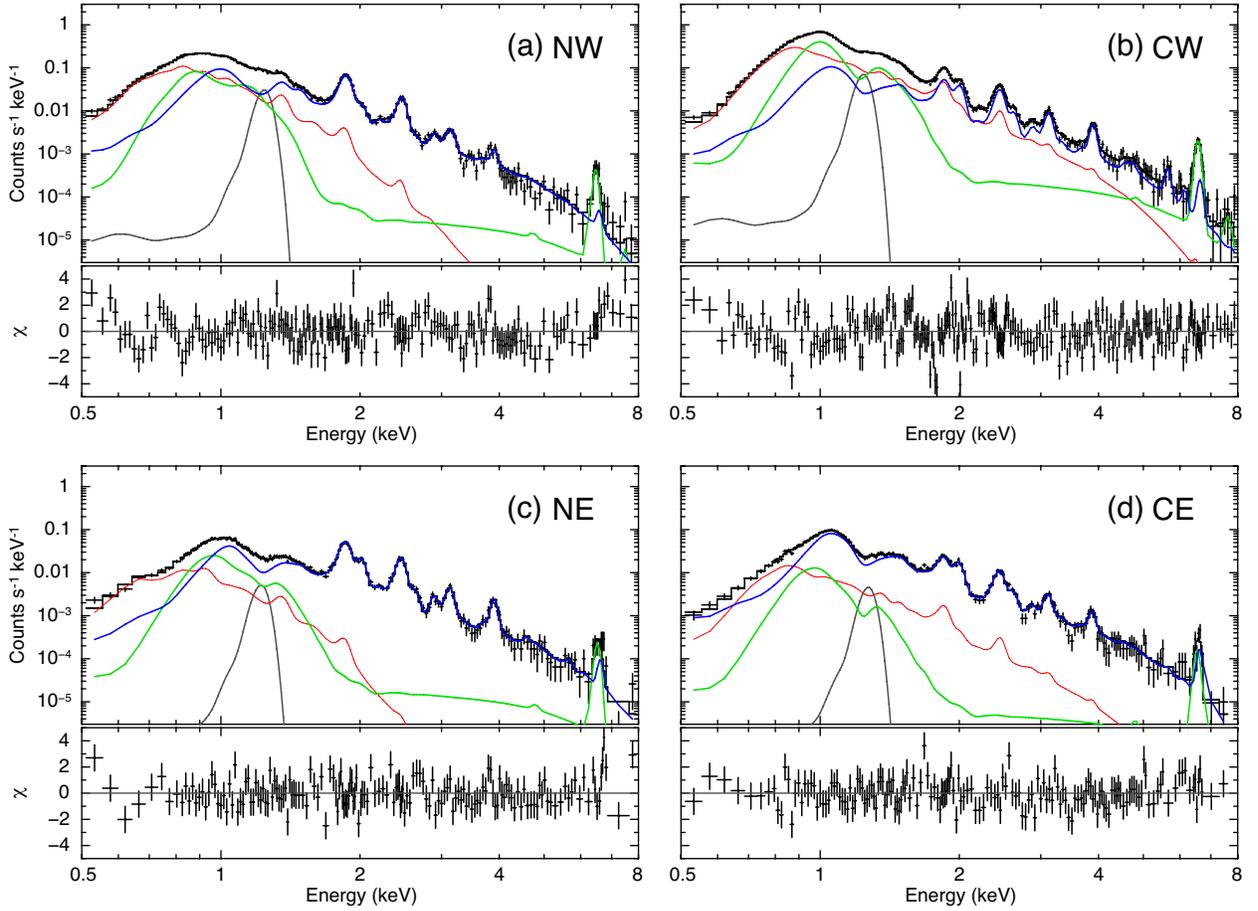}
	\caption{\footnotesize
		ACIS spectra of (a) NW, (b) CW, (c) NE, and (d) CE. 
		The best-fit models for the CSM, Fe ejecta, and IME ejecta components 
		are given as red (thin), green, and blue lines, respectively. 
		The gray curve is an additional Gaussian component 
		to compensate for incompleteness of the atomic data. 
		Spectra from different observations are merged and
		binned only for the display purpose. 
	\label{all_spectra}}
	\vspace{0.5mm}
  \end{center}
\end{figure*}

The model described above gives a good fit to the spectrum of each region 
in the 0.5--8.0\,keV band, except for the energies around 1.25\,keV. 
The discrepancy between the data and model found around this energy is likely 
due to the incompleteness of the atomic data (e.g., inaccurate emissivity of 
high-level L-shell transitions of Fe and/or Ni). 
We thus add a single Gaussian to compensate for this incompleteness. 
The best fit is then obtained as given in Figure~\ref{all_spectra} and Table~1. 
The results confirm the validity of our identification of the three components (\S2.2): 
The soft X-rays that dominate the O K emission are well reproduced by the low-\kTe\ plasma with 
the LMC mean abundances. The Fe K-shell emission and a part of the L-shell emission are 
successfully modeled with a pure-metal Fe-rich plasma. The emission from the IME requires 
another component with significantly enhanced abundances of these elements with respect to 
the LMC mean values. We also confirm that the ionization parameter ($\tau$) of the IME ejecta 
is highest at the CE region and lowest at the NW region, as inferred from Figure~\ref{mean_si}a. 
This result is in contrast to the previous work by \cite{Lewis03}, where little spatial 
variation in the ionization parameter was found for the IME component. 
\cite{Lewis03} also claimed that the IME ejecta component was virtually in the CIE. 
Our results, on the other hand, indicates the evidence for the NEI 
($\tau$ $\lesssim$ $10^{11}$\,cm$^{-3}$\,s). This discrepancy arises likely because 
the previous work used data from the early Chandra observations with much shorter 
exposure, and their spectral analysis focused solely on the radial trend assuming 
no azimuthal variation in the plasma properties.

The ionization parameter of the Fe ejecta is found to be significantly lower than 
that of the IME ejecta in any regions (Table~1). This is a common characteristic of 
SNRs Ia \citep[e.g.,][]{Badenes07,Sawada19,Fukushima20} and indicates the stratified 
elemental composition with Fe at the interior so that the Fe ejecta are heated 
by the reverse shock more recently than the IME ejecta. 
Such elemental stratification is indeed observed in the radial profile shown in Figure\,2.

We detect relatively strong Cr K emission at $\sim$\,5.6\,keV in the CW spectrum, 
which requires a high Cr abundance of the IME ejecta component ($61_{-11}^{+14}$~solar). 
Given that the CW region is dominated by the Fe ejecta component rather than 
the IME ejecta, the Cr in this region could be associated with the former. 
In fact, the spectrum can also be well reproduced if we 
fit the Cr abundance of the Fe ejecta component, 
instead of the assumption in our baseline model (Table~1). 
In this case, the Cr/Fe abundance ratio is obtained to be $\sim$\,3.6~solar.

The absorption column density of the NE region is found to be 
about twice higher than that of the other regions. 
We note, however, that the currently available data do not allow us 
to conclude whether this difference is real. 
Even if we fix the absorption column at $3.0 \times 10^{21}$\,cm$^{-2}$, 
a good fit is obtained with a c-stat/d.o.f.\ value of 3923/6127, 
a slightly higher \kTe\ for the CSM component ($0.23 \pm 0.01$\,keV), and a 
lower Mg abundance for the IME ejecta component ($0.60_{-0.08}^{+0.25}$~solar). 
The other spectral parameters do not change significantly from the best-fit values 
in Table~1.

\subsection{Charge Balance}
\label{ssec:charge}

\begin{figure}[t!]
  \begin{center}
	\includegraphics[width=7.8cm]{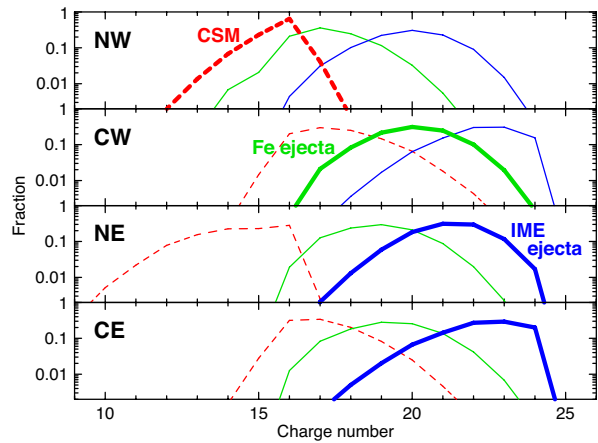}
	\caption{\footnotesize
		Charge distribution of Fe ions in each component in each region. 
		Red (dashed), green, and blue are the CSM, Fe ejecta, and IME ejecta components, 
		respectively. 
		The bold lines represent the dominant component in the spectra 
		as reported in Table~1. 
	\label{charge_dist}}
	\vspace{-0.4cm}
  \end{center}
\end{figure}

In Figure~\ref{charge_dist}, we show the charge distributions 
of Fe in each component in each region, calculated using the plasma parameters  
obtained in Table~1. We confirm that Fe ions in the IME component are more highly ionized 
than in the others, as inferred from the spatial correlation between the mean energy of 
the Fe L emission and the fraction of the IME component (Figure\,\ref{mean_fe}).

\begin{figure}[t!]
  \begin{center}
          \vspace{2mm}
	\includegraphics[width=7.8cm]{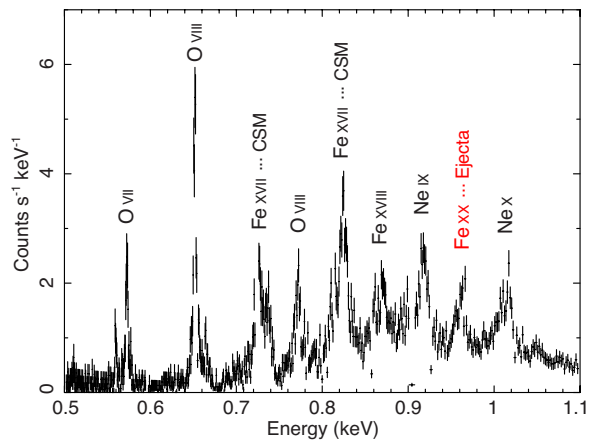}
	\caption{\footnotesize
		RGS1 first-order spectrum of N103B in 0.5--1.1\,keV, 
		extracted from the XMM-Newton archival data of ObsID = 0113000301.
	\label{spec_rgs}}
  \end{center}
\end{figure}

Figure~\ref{charge_dist} also indicates the difference in the Fe charge population between 
the CSM and Fe ejecta components; the former is dominated by Fe$^{16+}$ and Fe$^{17+}$, 
whereas the latter by Fe$^{19+}$ and Fe$^{20+}$. Notably, this result explains why N103B 
was once misinterpreted as a core-collapse SNR by the XMM--Newton/RGS study \citep{Heyden02}. 
To help the following discussion, we show in Figure~\ref{spec_rgs} 
the first-order RGS spectrum of the whole SNR obtained from the archival 
XMM-Newton data of ObsID = 0113000301, which exhibits the strong 
O\,{\footnotesize VII} and O\,{\footnotesize VIII} emission as well as 
the Fe\,{\footnotesize XVII} emission. 
In the previous work, it was assumed that those emission lines originate 
from the SN ejecta, and the near-solar Fe/O abundance ratio was obtained.
This result led to the core-collapse interpretation, since SN Ia nucleosynthesis 
models generally predict much higher Fe/O ratios. 
However, the O emission from this SNR is now known to be associated with 
the CSM \citep[e.g.,][and this work]{Williams14,Li17,Ghavamian17}, 
and our analysis indicates that the Fe\,{\footnotesize XVII} emission 
is also dominated by the the CSM component. 
We can therefore conclude that the previous measurement 
simply represents the elemental composition of the CSM, 
and is not in contradiction to the SN Ia classification.

\begin{figure}[t!]
  \begin{center}
          \vspace{2mm}
	\includegraphics[width=7.8cm]{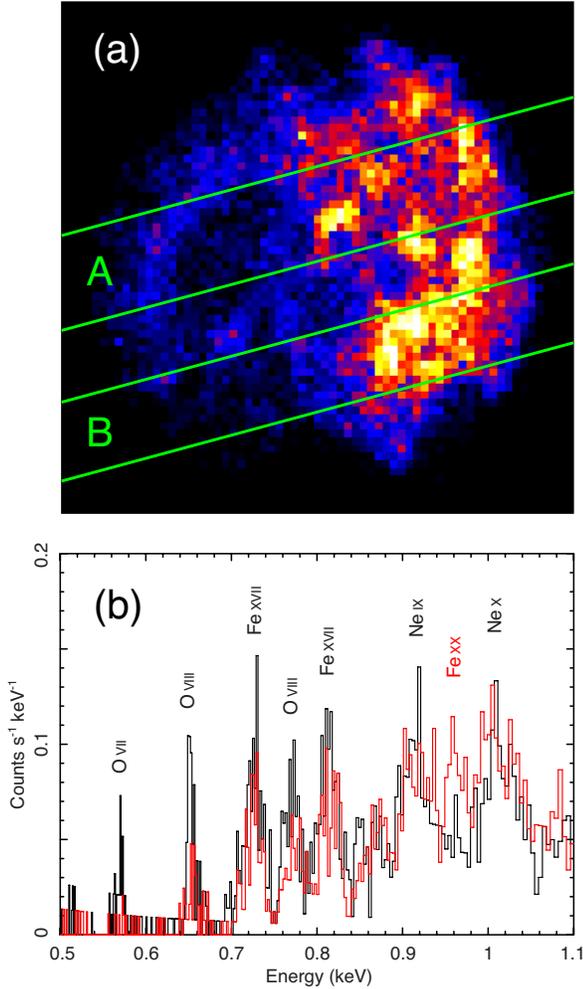}
	\caption{\footnotesize
		(a) Zeroth-order image of N103B from the archival data of 
		the Chandra HETG observations of ObsID = 1045. 
		The green lines indicate the dispersion direction of the MEG 
		as well as where the spectra of the CSM dominant region (A) 
		and the Fe ejecta dominant region (B) are extracted. \ 
		(b) Negative first order MEG spectra of the regions A (black) 
		and B (red). The emission at $\sim$\,0.96\,keV 
		is prominent only in the latter. 
	\label{spec_hetg}}
  \end{center}
\end{figure}

It is also worth noting that the Fe\,{\footnotesize XX} L-shell emission is detected in 
the RGS spectrum at $\sim$\,0.96\,keV, which likely originates from the Fe ejecta 
according to our analysis. This inference is further supported if a spatial 
correlation between this emission and the Fe K emission is revealed. 
Figure~\ref{spec_hetg}a shows a zeroth-order image from the archival data of Chandra 
High-Energy Transmission Grating (HETG) observations of N103B (ObsID = 1045), 
where the dispersion direction of the Medium Energy Grating (MEG) is indicated 
with the green lines. The regions A and B correspond to where the CSM and Fe ejecta 
components are dominant, respectively. 
Figure~\ref{spec_hetg}b shows negative first order spectra of the MEG from the regions 
A (black) and B (red), confirming that the Fe\,{\footnotesize XX} emission at 
$\sim$\,0.96\,keV is indeed prominent in the latter (i.e., Fe ejecta). 
Unfortunately, dispersive spectrometers like the RGS and HETG are not up to 
further detailed spatially resolved spectroscopy of extended sources. 
Future microcalorimeter observations with sufficient spatial resolution, 
which will be enabled by the Athena X-ray Integral Field Unit \citep[X-IFU:][]{Barret18}, 
are necessary for better identification of the major origin of each individual emission line. 
We also emphasize that, in principle, our component separation method (GMCA) could work 
even better for high spectral resolution data cubes as the spectral diversity of 
physical components would be enhanced.

\section{Interpretation and Discussion}
\label{sec:interpretation}

In this work, we have applied the GMCA technique \citep{Bobin16,Picquenot19} 
to the deep Chandra observations of N103B, discovering the double-ring structure 
of the IME ejecta as well as the more spatially confined Fe-rich ejecta of this SNR. 
The ionization degree of the IME ejecta is highest in the CE region 
that corresponds to the east edge of the west ring. 
This indicates that the CE region is the geometrical outermost of the SNR, 
although located near the projected SNR center. 
The spatial distribution of the IME ionization state 
(represented by the Si Ly$\alpha$/He$\alpha$ ratio) shows a clear bimodality, 
high/low in the East/West bubbles, suggesting that the two bubbles are 
spatially isolated from each other. 
We also find that the IME ejecta in the NE/CE regions 
(both located at the geometrical outermost) are significantly blue/redshifted 
with respect to the local LMC ISM ($v \sim 260$\,km\,s$^{-1}$). 
Although these measurements are subject to the uncertainties in gain calibrations, 
the observed velocities may represent the bulk motion of the outermost ejecta.

\begin{figure}[t!]
  \begin{center}
    \vspace{2mm}
	\includegraphics[width=8cm]{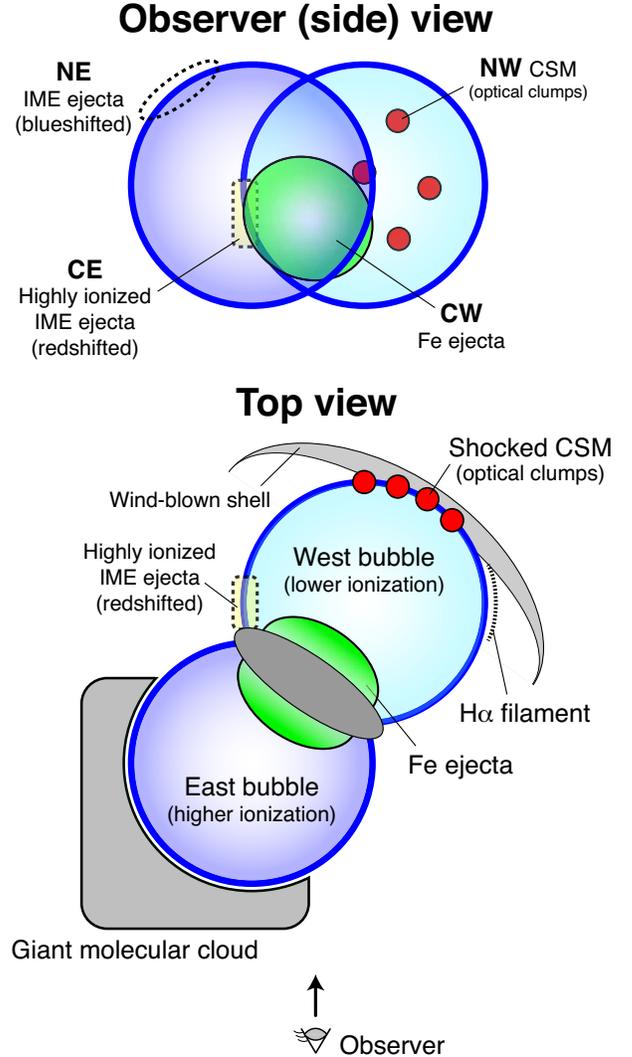}
	\caption{\footnotesize
		Schematic view of the interpreted geometry of the SNR N103B and  
		its environment.
	\label{schematic}}
  \end{center}
\end{figure}

As a coherent interpretation of these observational results and the previous 
multiwavelength studies of this SNR, we propose that N103B 
forms a bipolar structure, similarly to SN\,1987A  
and typical planetary nebulae (e.g., NGC\,2346), 
and is seen as a double-ring shape by the projection. 
A schematic of our interpretation is shown in Figure~\ref{schematic}. 
We assume that a dense CSM disk or torus (gray ellipse in the figure) had been 
formed around the progenitor prior to the SN explosion, causing a faster expansion 
of the SN ejecta toward the polar directions to form the bipolar bubbles. 
A similar geometry is suggested for Kepler's SNR, where a substantial amount of CSM is 
found around the projected SNR center (in addition to the more prominent northern rim), 
leading to an interpretation of disk distribution of the CSM at the equatorial plane 
of the progenitor binary system \citep{Burkey13,Chiotellis20}. 
An important difference between the two objects is that the system is viewed at 
an angle of $\sim$\,$45^{\circ}$ in N103B compared to nearly edge-on in Kepler's SNR, 
so the bipolar shells of N103B are partially overlapping with each other on 
the projected sky.

\begin{figure*}[t!]
  \begin{center}
        \vspace{1mm}
	\includegraphics[width=16.4cm]{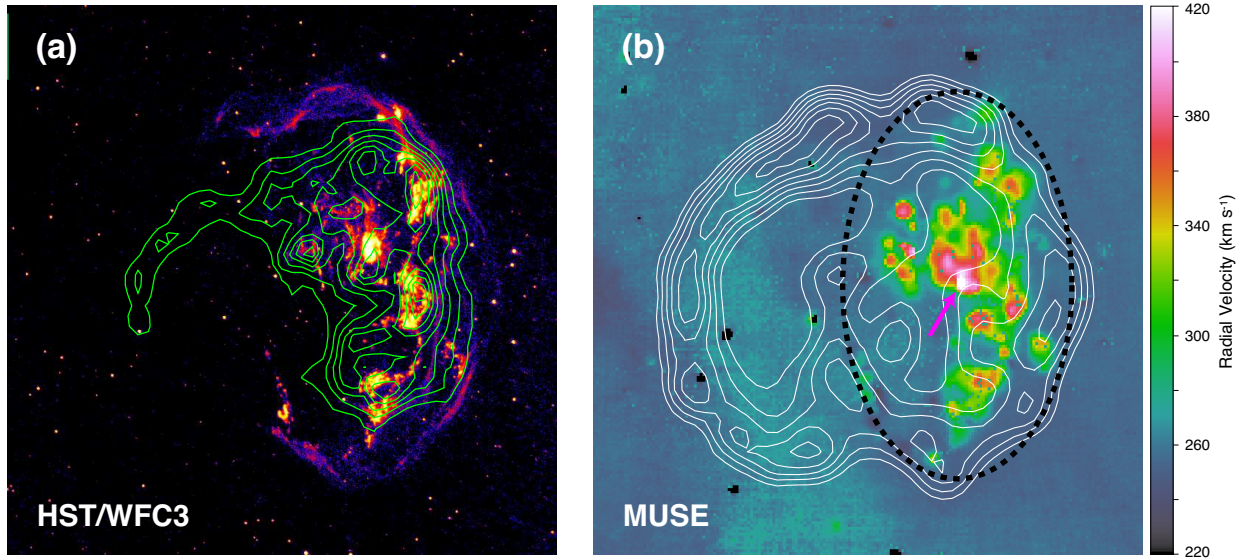}
	\vspace{2mm}
	\caption{\footnotesize
		(a) H$\alpha$ image of the SNR N103B with the HST/WFC3, 
		overplotted with the contours of the CSM component 
		(the same as Figure\,1b).
		Both are well correlated with each other. \ 
		(b) Radial velocity of the H$\alpha$ clumps measured with 
		the MUSE observation, overplotted with the contours of the IME ejecta 
		(the same as Figure\,3b). 
		The local ISM velocity ($\sim$\,260\,km\,s$^{-1}$) is not subtracted.
		The highest velocity is observed at the region pointed by the magenta arrow. 
		The black ellipse indicates the rough position of the H$\alpha$ filaments, 
		which corresponds to the ``west bubble''.  
		The identical sky region is presented in both panels.
	\label{hst}}
  \end{center}
\end{figure*}

The X-ray surface brightness of N103B is $\sim$\,5 times higher in the west than 
in the east, indicating that the current plasma density is about twice higher in 
the west (since the X-ray flux is proportional to $n_{\rm e}^2$). 
In contrast, the ionization degree of the IME ejecta, which depends on 
$\int n_{\rm e} \, dt$, is substantially higher in the east.  
This apparent discrepancy implies a complex density gradient in the pre-explosion 
ambient medium. 
The density immediately around the progenitor must have been 
higher toward the east, so the IMEs ejected to this direction got highly ionized 
shortly after the SN explosion. On the other hand, the west bubble must have 
expanded in a lower-density cavity during the early SNR evolution, and have 
recently reached a dense wind-blown shell.

Figure~\ref{hst}a shows an H$\alpha$ image of N103B obtained by the
Wide Field Camera 3 (WFC3) aboard the HST \citep{Li17}. 
The faint filaments found in the western half represent collisionless forward shocks 
propagating into the mostly neutral ambient medium in the cavity. 
The H$\alpha$ image also exhibits dense radiative clumps {\it inside} the 
Balmer-dominated filamentary shell, which coincide with the CSM component 
identified in X-rays (the green contours in Figure~\ref{hst}a). 
The H$\alpha$ emission from these clumps is significantly redshifted with 
respect to the local background and filamentary shell \citep{Li17,Ghavamian17}, 
suggesting that the clumps are physically located at the backside of the west bubble, 
where the SNR forward shock is interacting with the wind-blown shell. 
The total mass of the optical clumps is estimated to be 0.1--3\,\Msun\ 
\citep{Williams14,Li17,Blair20}.
Figure~\ref{hst}b presents the radial velocity of the H$\alpha$ clumps measured 
using the Multi Unit Spectroscopic Explorer (MUSE) on the Very Large Telescope. 
Details of the observations and data reduction will be presented in 
a separate paper (Li et al.\ in preparation). 
We reveal that the highest radial velocity is achieved at the middle of the west 
bubble (indicated as the magenta arrow in Figure~\ref{hst}b). This result supports 
our bipolar shell scenario; if the SNR is spherically symmetric with respect to 
the geometric center of the X-ray emission, a higher radial velocity should be 
achieved further east.

Interestingly, the eastern half of the SNR shell is completely missing 
in the H$\alpha$ image. To interpret the lack of this emission, previous 
optical/infrared studies suggested that the progenitor binary system had 
a high proper motion toward the west, creating an asymmetric distribution 
of the CSM \citep{Li17,Williams14}. 
This interpretation is analogous to Kepler's SNR, where the north--south 
contrast of the CSM density is indeed caused by the proper motion of the 
progenitor \citep{Blair07,Chiotellis12,Williams12,Burkey13}. 
However, recent radio observations with the ATCA and ALMA discovered 
a giant molecular cloud interacting with the southeast rim of the SNR \citep{Sano18}, 
whose average density ($n_{\rm H_2}$ $\sim$ 1500\,cm$^{-3}$) is even higher 
than that of the H$\alpha$ clumps ($n_{\rm H}$ $\sim$ 500\,cm$^{-3}$).
This suggests that the asymmetry of the H$\alpha$ emission is owing to 
the difference in the phase of hydrogen rather than the density contrast; 
the ambient medium to the east is dominated by molecules and thus 
hardly emits the Balmer lines.

Unlike the case of Kepler's SNR, we have not yet conclusively identified 
emission from the central CSM torus. We note, however, that the projection makes 
it difficult to spatially resolve the torus (if any) from the backside CSM in 
the bright western half. Figure\,\ref{image_comp}e shows that the relatively 
faint emission from the shocked CSM is present in the eastern half of the SNR. 
Moreover, the brightness profile of this component peaks at a smaller 
radius than those of the IME and Fe ejecta (Figure\,\ref{profile}). 
If this component is associated with the central CSM torus, 
a redshift of $\sim$\,$1.1~(v/500\,{\rm km\,s}^{-1})$~eV is expected 
for the O\,{\footnotesize VIII} lines, which can only be detected by 
high-resolution spectrometers with adequate angular resolution, 
such as the Athena X-IFU.

A torus-like CSM as well as bipolar cavities are often observed in planetary nebulae. 
In the case of N103B, however, the companion is suggested to be a main-sequence star 
from the abundance study \citep{Blair20}, which rules out a planetary nebula origin, 
or the so-called core-degenerate scenario \citep[e.g.,][]{Kashi11,Tsebrenko15,Chiotellis20}. 
In the context of the SD scenario, it is theoretically 
predicted that optically thick winds from the progenitor WD are driven by 
the mass accretion from the main-sequence companion \citep{Hachisu96}. 
When the winds are strong enough, they collide with the companion and strip off its 
surface layer \citep{Hachisu03b,Hachisu03a}. The stripped-off materials then form a 
massive (a few \Msun) circumstellar torus on the orbital plane \citep{Hachisu08a}, 
which leads to the formation of hourglass-like cavities toward the polar direction 
by the fast wind from the WD. 
Notably, this scenario suggests a relatively short delay time of $\sim$\,100\,Myr
(i.e., the age of the progenitor system at the SN Ia explosion), consistent with 
the estimate from the star formation history in the SNR site \citep{Badenes09}. 
The efficient optically thick wind requires high metallicity \citep[e.g.,][]{Kato94}, 
which is also consistent with another observational constraint for N103B 
\citep[$\gtrsim$\,$1\,Z_{\odot}$:][]{MR17}.

Interaction between SN ejecta and torus-like CSM
was suggested for SN 2012dn, a candidate of the bright, super-Chandrasekhar SN Ia 
\citep{Yamanaka16,Nagao18}. Kepler's SNR is also thought to originate from 
a luminous SN Ia \citep{Patnaude12,Katsuda15a}. 
While the peak luminosity of the supernova N103B has not been robustly determined 
due to the lack of published light echo spectroscopy, the presence of the strong 
Fe \Ka\ emission implies its luminous SN Ia origin \citep{Yamaguchi14b,MR18}. 
An association between SD progenitors and luminous SNe Ia is therefore suggested, 
in line with some theoretical predictions \citep[e.g.,][]{Fisher15}.


\section{Conclusions}
\label{sec:conclusions}

We have presented an X-ray study of N103B, 
a young SNR Ia that is known to be interacting with the dense ambient medium. 
Applying our novel GMCA method to the deep ($\sim$\,400-ks) Chandra observations, 
we have discovered the double ring structure of the IME ejecta. 
Our detailed spectroscopic study has allowed us to obtain a better 
understanding of the three-dimensional geometry of the SNR. 
The preferred scenario is that the pre-explosion stellar winds from the 
progenitor system had created a dense CSM torus 
and an hourglass-shape cavity, in which the SNR is expanding to form 
the bipolar shells. This scenario strongly favors an SD progenitor as 
the origin of this SNR, consistent with the previous multiwavelength study 
\citep[e.g.,][]{Williams14,Li17,Sano18}.

There are several open issues left for our future work.
First, it is crucial to identify X-ray emission from the central CSM torus 
to provide conclusive evidence of our bipolar geometry scenario. 
Although we have argued that the relatively faint CSM component extended to the 
east (see Figure\,\ref{image_comp}e) could be associated with the central torus, 
measurement of its radial velocity is necessary to conclude if this is indeed the case. 
Such study requires spatially-resolved high-resolution spectroscopy that will be 
enabled by the Athena X-IFU. 
Second, although it has not been the main focus of the present work, 
our GMCA has determined the detailed morphology of the Fe ejecta as well, 
revealing an intriguing ring structure around the SNR center (Figure\,\ref{image_comp}c). 
Further investigation is required to determine if this structure is formed 
due to the SN explosion itself or interaction with the ambient medium.


\acknowledgments

We thank Parviz Ghavamian, Hidetoshi Sano, Yasuo Fukui, Keiichi Maeda, 
Ken'ichi Nomoto, Brian Williams, and Benson Guest for helpful discussion about the 
previous multi-wavelength studies of the SNR N103B and/or the interpretation of 
the observational results. 
H.\,Y.\ is supported by Grants-in-Aid for Scientific Research (KAKENHI) of 
the Japanese Society for the Promotion of Science (JSPS) grant 
Nos.\ JP19H00704 and JP20H00175. 
C.-J.\,L.\ and Y.-H.\,C.\ are supported by the grants MOST 109-2112-M-001-040
and 109-2811-M-001-545 from the Ministry of Science and Technology of Taiwan.

\bigskip





\end{document}